\documentclass[twocolumn]{revtex4-2}
\usepackage[dvipdfmx]{graphicx}
\usepackage{amsmath,amsbsy,amssymb}
\usepackage{bm}
\usepackage{mathrsfs}
\usepackage{ulem}
\usepackage{textgreek}
\usepackage{mathrsfs}
\usepackage{mleftright}
\usepackage{braket}
\usepackage{physics}
\usepackage{hyperref}

\graphicspath{{./figures/}}

\usepackage{color}

\newcommand{\diff}{\mathrm{d}}
\newcommand{\imag}{\mathrm{Im}\,}
\newcommand{\imu}{\mathrm{i}}
\newcommand{\epn}{\mathrm{e}}
\newcommand{\sgn}{\mathrm{sgn}\,}
\newcommand{\ua}{\uparrow}
\newcommand{\da}{\downarrow}
\newcommand{\dg}{\dagger}
\newcommand{\la}{\langle}
\newcommand{\ra}{\rangle}

\newcommand{\sg}{\sigma}
\newcommand{\gm}{\gamma}
\newcommand{\ep}{\varepsilon}

\begin{document}

\title{
Eliashberg theory of the Jahn-Teller-Hubbard model
}

\author{
Yoshikuni Kaga$^1$, Philipp Werner$^2$, and Shintaro Hoshino$^1$
}

\affiliation{
$^1$Department of Physics, Saitama University, Saitama 338-8570, Japan
\\
$^2$Department of Physics, University of Fribourg, 1700 Fribourg, Switzerland
}

\date{\today}

\begin{abstract}
We study a multiorbital Hubbard model coupled to local Jahn-Teller phonons to investigate the superconducting state realized in fullerides.
A weak-coupling approach is employed in combination with a local self-energy approximation.
In addition to the normal and anomalous self-energies of the electrons, we  consider the phonon self-energy, which allows a self-consistent treatment of the energetics.
The frequency dependence of the self-energies and their characteristic coefficients, such as renormalization factors and dampings, are investigated in detail using numerical calculations.
It is clarified that the anisotropic phonons play an important role in the stabilization of the superconducting state.
By comparing the full 
results to those without phonon self-energies, we show that the superconductivity is stabilized by the softening of the phonon frequency.
The effects of electronic fluctuations are also considered, which leads to the coupling to orbitons, an analog of plasmons in the electron gas.
This additional contribution further stabilizes the superconducting state.
\end{abstract}

\maketitle

\section{Introduction}
Electron-phonon coupling is ubiquitous in solids, and it causes a variety of interesting phenomena.
One prominent example is superconductivity: the effective interaction between electrons becomes attractive, due to contributions mediated by the phonons, and results in Cooper pair formation.
Although in the conventional BCS theory the attractive interaction is treated as instantaneous, the actual electron-phonon interaction leads to a retardation effect in the effective interaction.
This situation is described by the Eliashberg theory \cite{Eliashberg60, Eliashberg63}, which allows to incorporate the retardation effect.
Recently, band-structure calculations have also been performed within this framework, and they have shown success in predicting the superconducting transition temperature \cite{Akashi15,Flores-Livas16,Sano15,Sanna18,Wang20,Ikhsanov22}.

Usually, the phonons 
describe {\it atomic} oscillators
in solids.
The situation is however different in the fulleride superconductors \cite{Hebard91,Rosseinsky91,Holczer91,Tanigaki91,Fleming91,Ganin08,Takabayashi09,Mitrano16,Ren20}.
Since these are 
molecular crystals,
it is important to consider the physics of each molecular unit.
In fact, in these materials, the dominant phonon contribution 
is associated with 
{\it local molecular} vibrations.
As for the electron-electron interaction, the Coulomb interaction inside the molecule also plays an important role. 
Thus, the most relevant interactions have a local nature. 
The simplest model to describe such a situation is the Holstein-Hubbard model, where a single orbital is assumed and the electron charge couples to spatially local Einstein phonons.
However, the existing fulleride superconductors have three degenerate $t_{1u}$ (or $t_{1g}$) molecular orbitals, and the electrons in this multi-orbital system
couple to spatially local anisotropic molecular vibrations (Jahn-Teller phonons) \cite{Gunnarsson97,Capone09,Nomura16}.
We therefore need to analyze the Jahn-Teller-Hubbard model with multiple electron/phonon degrees of freedom in order to understand the physics of fullerides.

In the three-dimensional fullerides with cubic symmetry, intersite correlations are relatively weak so that the dynamical mean-field theory (DMFT), which takes into account local correlations, is a suitable theoretical approach \cite{Georges96}.
Due to the coupling to  Jahn-Teller phonons, the effective Hund's coupling for the $t_{1u}$ orbitals becomes antiferromagnetic at low energies \cite{Fabrizio97}.
For this reason, multiorbital Hubbard models with antiferromagnetic Hund's coupling have been intensively investigated \cite{Capone02,Capone04,Hoshino16,Steiner16,Hoshino17,Ishigaki18,Ishigaki19,Hoshino19,Yue20,Yue21}.
For a more accurate treatment, however, we need to explicitly consider the phonon degrees of freedom and the associated retardation effects.
In the fulleride context, systems with both electron-electron and electron-phonon interactions have been analyzed using quantum Monte Carlo methods \cite{Han00, Han03, Nomura15, Steiner15} and weak-coupling perturbation theory \cite{Yamazaki13,Yamazaki14}.
However, since the interactions between electrons and phonons in multi-orbital systems are quite complicated, they have been simplified in these theoretical studies.

In this paper, we analyze the model by using the Eliashberg theory within the framework of DMFT, which provides an intuitive understanding of the Jahn-Teller Hubbard model.
We focus on the fact that the electron-electron interaction and electron-phonon interaction can be handled in a unified way by introducing the charge-orbital (multipolar) moments \cite{Iimura21}.
This makes it easier and more transparent to formulate the Eliashberg equations for the Jahn-Teller Hubbard model.
While the multiorbital nature complicates the theoretical formulation, the local nature of the correlation effects 
allows to reveal
the structure of the self-consistent equations.
We will also analyze the model in an analytic way in the normal state, which provides some intuitive understanding of the underlying physics.
A more elaborate analysis will be done with the help of numerical calculations.
Based on this theoretical approach, we clarify how the superconducting transition temperature is determined.
Specifically, we discuss the effect of the phonon self-energies and also propose a possible further improvement of the theory by introducing the `orbiton', which is an analog of the plasmon in the theory of the electron gas.

This paper is organized as follows.
In the next section, we derive the self-consistent Eliashberg equations from the variational principle.
In Sec.~III, the phonon self-energy is introduced, which is necessary for the self-consistent treatment of the internal energy.
Section~IV is devoted to the analysis of the normal Fermi liquid in the weak-coupling limit.
We provide detailed numerical results in Sec.~V.
In Sec.~VI, we study a more complicated situation by considering the fluctuations from the electron-electron interaction.
Finally, we summarize the results in Sec.~VII.
Appendix A explains the connection to the actual C$_{60}$ molecule.
Convenient formulae and some calculation details are provided in Appendices B and C.

\section{Self-consistent equations}

\subsection{Jahn-Teller-Hubbard model}

We consider the three-orbital Hubbard model coupled to isotropic and anisotropic phonons.
The Hamiltonian is written as
\begin{align}
    \mathscr H &= \mathscr H_{e} + \mathscr H_{p} + \mathscr H_{ep} 
    .
\end{align}
The electron part is given by $\mathscr H_e = \mathscr H_{e0} + \mathscr H_{\rm int}$, where $\mathscr H_{e0}$ describes the non-interacting part and $\mathscr H_{\rm int}$ is the local Coulomb interaction with the Slater-Kanamori form 
\begin{align}
    \mathscr H_{\rm int} &= \frac 1 2 \sum_{i \sg\sg'}\sum_{\gm_1\gm_2\gm_3\gm_4}
    U_{\gm_1\gm_2\gm_3\gm_4} c^\dg_{i\gm_1\sg} c^\dg_{\gm_2\sg'} c_{\gm_4\sg'} c_{\gm_3\sg},
\end{align}
where $i$ is the lattice site index and $\sg=\ua,\da$ denotes spin.
The orbital index is written as $\gm = x,y,z$ 
and refers to the 
molecular $t_{1u}$ orbitals relevant for alkali-doped fullerides.
We use the standard parametrization for the interaction: $U_{\gm\gm\gm\gm} = U$ is the intra-orbital interaction, $U_{\gm\gm'\gm'\gm} = U'$ is the inter-orbital interaction, and $U_{\gm\gm'\gm\gm'}=U_{\gm\gm\gm'\gm'} = J$ ($>0$) is the ferromagnetic Hund's coupling ($\gm\neq\gm'$).
Assuming spherical symmetry in the interaction, we set $U'=U-2J$.
The non-interacting part will be specified later.

In the phonon parts, we consider local phonons arising from the dynamical deformation of the fullerene molecule, which couples to the orbital degrees of freedom of the electrons.
The Hamiltonian is given by
\begin{align}
    \mathscr H_{ep} &= \sum_{i\eta} g_\eta \phi_{i\eta} T_{i\eta},
    \label{eq:ham_ep}
    \\
    \mathscr H_p &= \sum_{i\eta}\omega_\eta a^\dg_{i\eta} a_{i\eta},
\end{align}
where $\phi_{i\eta} = a_{i\eta} + a_{i\eta}^\dg$ is a displacement operator for molecular vibrations.
The charge-orbital moment $T_{i\eta}$ ($\eta=0,1,3,4,6,8$) is introduced together with the Gell-Mann matrices as
\begin{align}
&T_{i\eta} = \sum_{\gm\gm'\sg}c^\dg_{i\gm\sg} \lambda^\eta_{\gm\gm'} c_{i\gm'\sg},
\\
&\hat \lambda^0 = \sqrt{\frac 2 3} 
\begin{pmatrix}
1 & 0 & 0 \\
0 & 1 & 0 \\
0 & 0 & 1 
\end{pmatrix}
,\ 
\hat \lambda^1 =
\begin{pmatrix}
0 & 1 & 0 \\
1 & 0 & 0 \\
0 & 0 & 0 
\end{pmatrix},
\nonumber \\
&\hat \lambda^3 =  
\begin{pmatrix}
1 & 0 & 0 \\
0 & -1 & 0 \\
0 & 0 & 0 
\end{pmatrix}
,\ 
\hat \lambda^4 =
\begin{pmatrix}
0 & 0 & 1 \\
0 & 0 & 0 \\
1 & 0 & 0 
\end{pmatrix},
\nonumber \\
&\hat \lambda^6 = 
\begin{pmatrix}
0 & 0 & 0 \\
0 & 0 & 1 \\
0 & 1 & 0 
\end{pmatrix}
,\ 
\hat \lambda^8 =\sqrt{\frac 1 3} 
\begin{pmatrix}
1 & 0 & 0 \\
0 & 1 & 0 \\
0 & 0 & -2 
\end{pmatrix},
\label{eq:def_Gell-Mann}
\end{align}
and describes the change in the isotropic ($\eta=0$) and anisotropic ($\eta=1,3,4,6,8$) electron charge distribution.
The hat symbol ($\hat \ $) indicates a $3\times 3$ matrix in the orbital space.
There are useful 
orthogonality
relations: ${\rm Tr\,}\hat \lambda^{\eta} \hat \lambda^{\eta'} = 2 \delta_{\eta\eta'}$, $(\hat \lambda^0)^2 = \frac 2 3 \hat 1$, $\sum_{\eta=8,3}(\hat \lambda^\eta)^2 = \frac{4}{3}\hat 1$ and $\sum_{\eta=1,6,4}(\hat \lambda^\eta)^2 = 2\, \hat 1$.
In cubic symmetric systems, such as fcc fullerides,  polynomial representations can be assigned to each component: $\eta = 0 \leftrightarrow r^2$ ($A_{1g}$), $\eta=8,3\leftrightarrow 3z^2-r^2, x^2-y^2$ ($E_g$), and $\eta=1,6,4 \leftrightarrow xy, yz, zx$ ($T_{2g}$), where $r^2=x^2+y^2+z^2$ and the irreducible representations are written in the parenthesis.
We note that the decomposition into irreducible representations relevant for the electronic degrees of freedom is $[t_{1u}\times t_{1u}] = A_{1g} + E_g + T_{2g}$ where $[\cdots]$ stands for the symmetric product corresponding to electric (time-reversal-even) degrees of freedom such as charge and orbital moments.

In an actual isolated fullerene molecule (point group $I_h$), there are 174 vibrational modes, and two $A_g$ and eight $H_g$ modes are coupled to the charge-orbital moments defined in the $t_{1u}$ electronic orbital space \cite{Gunnarsson97}.
This complicated situation is approximated by choosing one dominant contribution for each irreducible representation.
A more detailed discussion is provided in Appendix A.

Using the above Gell-Mann matrices, the Slater-Kanamori interaction can be rewritten in a  symmetric and compact form \cite{Iimura21},
\begin{align}
    \mathscr H_{\rm int} &= \sum_{i\eta} I_\eta :T_{i\eta} T_{i\eta}: ,
    \label{eq:int_rewrite}
\end{align}
where the colon (:) symbol represents the normal ordering.
The interaction parameters are $I_0 =\frac{3}{4}U - J$ and $I_1=I_3=I_4=I_6=I_8 =\frac{J}{2} $, as derived from the original $\mathscr H_{\rm int} $ and reflecting the spherical symmetry. 
One can recognize that the forms of the electron-electron interaction in Eq.~\eqref{eq:int_rewrite} and the electron-phonon interaction in Eq.~\eqref{eq:ham_ep} are similar, which makes the formulation transparent as will be shown in the following sections.
We can utilize this property to introduce the `orbiton', a bosonic excitation
associated with
the Slater-Kanamori interaction, as discussed in Sec. VI.
If one wants to describe the interaction in 
a cubic environment,
one has to use different values for 
$I_3$ ($= I_8$, $E_g$) and $I_1$ ($= I_4=I_6$, $T_{2g}$),
which corresponds to $U'\neq U-2J$ \cite{Iimura21}.
The interaction parameters are related to the Slater-Kanamori parameters by 
$I_0 = \frac{1}{4}(U+2U')$, 
$I_3 = \frac{1}{2}J$, and 
$I_1 = \frac{1}{4}(U-U')$, where $I_1 - I_3 = \frac{1}{4} (U'+2J - U)$ controls the cubic anisotropy.
In the following, we assume $I_1=I_3$.

It is convenient to trace out the phonons by employing the path-integral formalism.
In order to do this, we consider the partition function
\begin{align}
    Z &= \int \mathscr D c \mathscr D a \ \exp \big(- \mathscr S_e[c] - \mathscr S_{ep}[c,a] - \mathscr S_p[a] \big), 
\end{align}
where $c$ and $a$ are Grassmann and complex numbers associated with the electrons (fermions) and phonons (bosons), respectively.
The bosons can be integrated out, and the resultant partition function is given by
\begin{align}
    Z &= A \int \mathscr D c \ \exp \big(- \mathscr S_{\rm eff}[c] \big) ,
\end{align}
where $A$ is an unimportant constant.
The effective action is given by $\mathscr S_{\rm eff} = \mathscr S_{e0} + \mathscr S_{\rm int}$ with
\begin{align}
\mathscr S_{e0} &= \int \diff \tau \bar c(\tau) \partial_\tau c(\tau) + \int \diff \tau \mathscr H_{e0} (\tau),
\label{eq:free_action}
\\
\mathscr S_{\rm int } &=  \int \diff \tau \diff \tau'
\sum_{i\eta} U_\eta(\tau-\tau') T_{i\eta}(\tau) T_{i\eta} (\tau').
\end{align}
For simplicity, we have used a vector notation for the Berry phase term in Eq.~\eqref{eq:free_action}.
The retarded interaction is given by
\begin{align}
    U_\eta(\tau) &= I_\eta \delta(\tau) + g^2_\eta \mathscr{G}_\eta (\tau),
\end{align}
where
$\mathscr{G}_\eta(\tau) = - \la \mathcal T a_\eta(\tau) \bar a_\eta(0) \ra_0$ 
is the non-interacting boson Green function.
Here $\mathcal T$ indicates the imaginary-time ordering operator.
The explicit expression is
\begin{align}
\mathscr G_\eta (\tau) &= - \frac{\epn^{(\beta - \tau)\omega_\eta} \, \theta(\tau) + \epn^{-\tau \omega_\eta} \, \theta(-\tau) }{\epn^{\beta \omega_\eta}-1}.
\end{align}
We note that, while the formulations are written in terms of continuous imaginary time for simplicity, the actual calculation must be performed for discrete variables \cite{Negele_book}.

\subsection{Variational principle for the electrons}

The DMFT approximation, which allows to treat strong local correlation effects, is a suitable theoretical framework for  three-dimensional electron systems.
However, taking full account of the retarded interaction in multiorbital systems is technically challenging, and some additional approximations are needed.
Here, we take a different approach and analyze the above model 
by employing the variational principle.
We introduce the model action in the form
\begin{align}
&\mathscr S_{\rm model} = \mathscr S_{e0} +
\int \diff\tau\diff\tau' \sum_{i\gm\gm' \sigma\sigma'}\Big\{
\rho_{\sigma\sigma'}^{\gm\gm'}(\tau,\tau')\bar c_{i\gm\sigma}(\tau)c_{i\gm'\sigma'}(\tau') \nonumber \\
&\hspace{-2mm}
+\varDelta_{\sigma\sigma'}^{\gm\gm'}(\tau,\tau')\bar c_{i\gm\sigma}(\tau) \bar c_{i\gm'\sigma'}(\tau')
+\varDelta_{\sigma\sigma'}^{\gm\gm'*} (\tau,\tau') c_{i\gm'\sigma'}(\tau') c_{i\gm\sigma}(\tau)
\Big\},
\label{eq:aux_field_def}
\end{align}
where $\rho$ and $\varDelta$ are complex local auxiliary fields to be determined.
This model action is the most generic one with bilinear terms with respect to the Grassmann numbers.

At this point, we also introduce the single-particle Green function, which is defined by
\begin{align}
    \check G_{\gm\gm'}(ij;\tau-\tau') &= - \la \mathcal T
    \bm \psi_{i\gm}(\tau) \bar{\bm \psi}_{j\gm'}(\tau')
    \ra
    \\
    &\hspace{-10mm} = \begin{pmatrix}
    G_{\gm\gm'} (ij;\tau - \tau') & F_{\gm\gm'} (ij;\tau - \tau') \\
    \bar F_{\gm\gm'} (ij;\tau - \tau') & \bar G_{\gm\gm'} (ij;\tau - \tau') 
    \end{pmatrix},
\end{align}
where 
the quantum statistical average is defined with the model action as $\la \cdots \ra = ( \int \mathscr D c \cdots \epn^{-\mathscr S_{\rm model}} ) / \int \mathscr D c \,\epn^{-\mathscr S_{\rm model}}$.
The check symbol ($\check{\phantom{a}}$) indicates a $2\times 2$ matrix in the Nambu space.
The Nambu spinor has been introduced as 
$   \bm \psi_{i\gm} = (c_{i\gm\ua}, \bar c_{i\gm\da})^{\rm T}$ and $   \bar {\bm \psi}_{i\gm} = (\bar c_{i\gm\ua}, c_{i\gm\da})$.

We employ the Gibbs-Bogoliubov inequality for the variational free-energy functional $\mathscr F_{\rm var}$ given by
\begin{align}
    - \ln Z \leq \mathscr F_{\rm var} = - \ln Z_{\rm model}  + \la \mathscr S_{\rm eff} - \mathscr S_{\rm model} \ra,
\end{align}
where the partition function is $Z_{\rm model} = A\int \mathscr D c \, \epn^{-\mathscr S_{\rm model}}$.
Minimizing $\mathscr F_{\rm var}$, we obtain the self-consistent equation
\begin{widetext}
\begin{align}
\begin{pmatrix}
\rho_{\uparrow \uparrow}^{\gm\gm'}(\tau,\tau') & 2\varDelta_{\uparrow \downarrow}^{\gm\gm'}(\tau,\tau') \\
2{\varDelta_{\uparrow \downarrow}^{\gm'\gm}(\tau',\tau)}^\ast & -\rho_{\downarrow \downarrow}^{\gm'\gm}(\tau',\tau)
\end{pmatrix}    
=
2 \sum_{{\eta} \gm_{1}\gm_{2}} \mathscr{U}_{\eta}(\tau-\tau')
\begin{pmatrix}
     - \lambda_{\gm \gm_{2}}^{\eta}  \lambda_{\gm_{1} \gm'}^{\eta} G_{\gm_{2} \gm_{1}}(\tau-\tau')
    & 
     \lambda_{\gm \gm_{1}}^{\eta}  \lambda_{\gm' \gm_{2}}^{\eta}  F_{\gm_{1} \gm_{2}}(\tau-\tau') 
    \\[2mm]
     \lambda_{\gm_{2} \gm}^{\eta}  \lambda_{\gm_{1} \gm'}^{\eta} F_{\gm_{1} \gm_{2}}(\tau-\tau')
    & 
    - \lambda_{\gm' \gm_{2}}^{\eta}  \lambda_{\gm_{1} \gm}^{\eta} \bar G_{\gm_{1} \gm_{2}}(\tau-\tau')
\end{pmatrix}
, \label{eq:Eliashberg}
\end{align}
\end{widetext}
where we have defined the symmetrized interaction as
\begin{align}
    2\mathscr U_\eta(\tau-\tau')
    &= U_\eta(\tau-\tau') + U_\eta(\tau'-\tau)
    \label{eq:def_symmetric_U}
\end{align}
and defined $\hat G_{\gm\gm'}(\tau) \equiv \hat G_{\gm\gm'} (ii;\tau)$ assuming translational symmetry.
We note that the Hartree term just gives a chemical potential shift for a cubic-symmetric system and is neglected.
We have assumed  spin-singlet pairing and even imaginary-time dependence of $\varDelta_{\uparrow \downarrow}(\tau , \tau')$.

Next, we define the self-energy by
\begin{align}
    \mathscr S_{\rm model} &= \mathscr S_{e0} + \int \diff\tau\diff\tau' \sum_{i\gm\gm'}
    \bar {\bm \psi}_{i\gm} (\tau) \check \Sigma_{\gm\gm'}(\tau-\tau') \bm \psi_{i\gm'}(\tau').
\end{align}
Note that $\Sigma$ is introduced in the restricted local Nambu basis for a spin-singlet pair, while the auxiliary fields in Eq.~\eqref{eq:aux_field_def} can describe a more general situation.
The self-consistent equation can then be written in a simple matrix form,
\begin{align}
&\begin{pmatrix}
\hat \Sigma_{11}(\tau) & \hat \Sigma_{12}(\tau)\\
\hat \Sigma_{21}(\tau)  & \hat \Sigma_{22}(\tau)
\end{pmatrix}  
= \begin{pmatrix}
\hat \rho_{\uparrow \uparrow}(\tau,0) & 2\hat \varDelta_{\uparrow \downarrow}(\tau,0) \\
2{\hat \varDelta_{\uparrow \downarrow}^{\rm T}(0,\tau)}^\ast & -\hat \rho_{\downarrow \downarrow}^{\rm T}(0,\tau)
\end{pmatrix}   
\nonumber \\
&=
\displaystyle -2\sum_{{\eta} }\mathscr{U}_{\eta}(\tau)
\begin{pmatrix}
 \hat \lambda^{\eta}    & 0 \\
 0 &  -\hat \lambda^{\eta} 
\end{pmatrix}
\check G(\tau)
\begin{pmatrix}
\hat \lambda^{\eta}    & 0 \\
 0 &  -\hat \lambda^{\eta} 
\end{pmatrix},
\label{eq:scf}
\end{align}
where the diagonal and offdiagonal parts correspond to the normal and anomalous self-energies, respectively.
This Eliashberg equation is consistent with the weak-coupling perturbtation theory with respect to the electron-phonon coupling constant $g_\eta$ and the electron-electron interaction $I_\eta$, which confirms the validity of our equation.

Let us make a comment on the difference between our formulation and that utilizing the Stratonovich-Hubbard transformation.
In the latter case, we must be careful about the sign of the interaction term and the convergence of the Gaussian integrals.
Furthermore, the decoupling of the interaction using the auxiliary bosonic field is not unique \cite{Castellani79}.
With the variational principle used in the present paper, on the other hand, the equations are uniquely defined once the variational action is written down.
The symmetry of the original system is also automatically encoded in the effective action.

Models with retarded interactions have been numerically studied using DMFT combined with the continuous-time quantum Monte Carlo method \cite{Nomura15,Steiner15}.
For the three-orbital case, the interactions need to be truncated to density-density terms in the actual calculation because of the fermion sign problem \cite{Nomura15,Steiner15}.
While the electronic correlation effects in higher orders are incorporated, the symmetry is lowered from the original one in this approximation.
On the other hand, while only the lowest-order contributions are considered in our method, we include all the interaction terms and the full symmetry of the original Hamiltonian is preserved.

\subsection{Cubic-symmetric solution}

The cubic symmetry  simplifies the self-consistent equations considerably.
Within the restricted $t_{1u}$ orbital space, the cubic symmetry requires that the self-energy must be proportional to the identity matrix in orbital space.
Hence we can write the self-energy as
\begin{align}
    & \hat \Sigma_{11}(\imu \omega_n) = \hat \Sigma_{22}(\imu \omega_n) = \Sigma(\imu\omega_n) \hat 1,
    \\
    & \hat \Sigma_{12}(\imu \omega_n) = \hat \Sigma_{21}(\imu \omega_n) = \Delta(\imu\omega_n) \hat 1,
\end{align}
where we have chosen the phase of the anomalous self-energy as real.
(We write these relations with the fermionic Matsubara frequency $\omega_n = (2n+1)\pi T$ ($n \in \mathbb Z$) at temperature $T$.)
Correspondingly, the Green functions are also proportional to the identity matrix: $\hat G = G \hat 1$ and $\hat F = F\hat 1$.

With this, we obtain a much simplified Eliashberg equation:
\begin{align}
    &\Sigma(\tau) = - U_{\rm eff}(\tau) G(\tau),
    \label{eq:simplified_sigma}
    \\
    &\Delta(\tau) = U_{\rm eff}(\tau) F(\tau),
    \label{eq:simplified_delta}
    \\
    &U_{\rm eff}(\tau) =\frac{4}{3}\sum_\eta \mathscr U_\eta(\tau) .
    \label{eq:def_eff_int}
\end{align}
The last expression shows that each 
$\eta$ equally contributes to the effective interaction among the electrons.
This point will be discussed in Sec.~IV in more detail.
We note that we do not consider the possibility of orbital symmetry breaking, even though such a symmetry breaking can occur in the strongly correlated regime \cite{Hoshino16, Hoshino17, Ishigaki19, Werner17, Misawa17, Iwazaki21}.

The Dyson equation, which has now also a simple form due to the cubic symmetry, connects the local self-energy and Green function as
\begin{align}
    \begin{pmatrix}
    G(\imu \omega_n) \\
    F(\imu \omega_n)
    \end{pmatrix}
    &= 
    \int \diff \ep  \,
    \frac{\rho(\ep)}{[\imu \omega_n - \Sigma(\imu \omega_n)]^2 - \Delta(\imu\omega_n)^2 - \ep^2 }
    \nonumber \\
    &\ \ \ \times
    \begin{pmatrix}
    \imu\omega_n -\Sigma(\imu \omega_n)  +\ep\\
    \Delta(\imu \omega_n)
    \end{pmatrix}
    ,
\end{align}
where the intersite information is included in the density of states $\rho(\ep)$. 
We will take a semi-circular shape for $\rho(\ep)$, since we want to understand the qualitative features which do not depend on the details of the density of states.

\section{Phonon Self-energy}

As derived in the last section, the simple variational principle for the effective electron system results in a coupling to the free bosons.
Here, we consider the self-energy for the phonons and show that it is important for a self-consistent energetics in electron-phonon coupled systems.

In the presence of the self-energy, the $\la a^\dg a^\dg\ra$-type average becomes finite.
Hence, a representation involving $\phi_{\imu\eta} =a_{i\eta} + a_{i\eta}^\dg$ instead of $a_{i\eta}$ is more appropriate.
The corresponding Green function is defined by
\begin{align}
    D_\eta(\tau) &= - \la \mathcal T \phi_{i\eta}(\tau) \phi_{i\eta} \ra .
\end{align}
The relations among the phonon Green functions is detailed in Appendix B.
The off-diagonal part with respect to $\eta$ is zero because of the orthogonality theorem in the group theory under the cubic symmetry.
The Fourier transform is given by
\begin{align}
    D_\eta (\imu\nu_m)^{-1} &=D_{\eta 0} (\imu\nu_m)^{-1} - \Pi_\eta(\imu\nu_m),
    \\
    D_{\eta 0}(\imu\nu_m) &= \frac{2 \omega_\eta}{(\imu\nu_m)^2 -\omega_\eta^2},
\end{align}
where $\nu_m = 2m\pi T$ ($m\in \mathbb Z$) is the bosonic Matsubara frequency at temperare $T$.
We have assumed a local self-energy also for the phonons.
Note that the Dyson equation for the phonon part does not involve any intersite information, because of the local nature of the non-interacting Hamiltonian,  in contrast to the electron case.

The self-energy is obtained by weak-coupling perturbation theory.
Up to second order in the electron-phonon coupling $g_\eta$, we obtain
\begin{align}
\Pi_\eta(\imu\nu_m) &= g_\eta^2 \chi_\eta (\imu\nu_m),
\label{eq:phon_self_def}
\end{align}
where we define the charge-orbital correlation function
\begin{align}
\chi_\eta (\tau) &= - \la \mathcal T T_{i\eta}(\tau) T_{i\eta} \ra_{\rm conn}
\\
&= 4\big[ G(\tau) G(-\tau) - F(\tau)^2 \big]
\equiv \chi(\tau),
\label{eq:def_orb_corr}
\end{align}
and `conn' represents the contribution from the connected diagrams.
The phase of the pair potential, and therefore the pair amplitude (= anomalous Green function), is fixed as real.
The factor of $4$ originates from spin and the trace of the square of the Gell-Mann matrices.
Note that the charge-orbital correlation function $\chi_\eta$ is expressed only in terms of the electronic degrees of freedom.
It is not dependent on the index $\eta$ because vertex corrections are neglected.

Correspondingly, the effective interaction in Eq.~\eqref{eq:def_symmetric_U} is replaced by 
\begin{align}
    \mathscr U_\eta(\imu\nu_m) &= I_\eta + 
    \frac 1 2 
    g_\eta^2 D_\eta (\imu\nu_m),
    \label{eq:modified_effective_interaction}
\end{align}
which now includes the self-energy.
This may be derived by  second-order perturbation theory or by repeating the procedure in the last section with the phonon self-energy.

Although the phonon self-energy from the coupling to the electrons is sometimes neglected in 
strong-coupling theories \cite{Parks_book}, it is necessary for an accurate evaluation of the internal energy \cite{Wada64}.
We note that the phonon self-energy may be dropped in the free-energy {\it difference} between the normal and superconducting states \cite{Bardeen64}, but it should be kept in the free energy itself.
In the context of the Jahn-Teller-Hubbard model, the interaction energy can be explicitly written in terms of the self-energies as \cite{Wada64}
\begin{align}
    &2\la \mathscr H_{\rm int} \ra + \la \mathscr H_{ep} \ra
    \nonumber \\
    &= 6 T\sum_n \big[ \Sigma(\imu\omega_n) G(\imu\omega_n) + \Delta(\imu\omega_n) F(\imu\omega_n) \big]\epn^{\imu\omega_n 0^+},
    \label{eq:energy_elec} \\
    &\la \mathscr H_{ep} \ra 
    = - \sum_\eta T\sum_{m} \Pi_\eta(\imu\nu_m) D_\eta(\imu\nu_m) \epn^{\imu \nu_m 0^+}.
    \label{eq:energy_phon}
\end{align}
When the electron-electron interaction is absent, the two expressions must give the same result, and hence
the phonon self-energy is necessary for a self-consistent theory.
The equations obtained above ensure this self-consistency and can be used to derive the expression for the phonon self-energy.

Practically, if we consider only the electron self-energies, the electron-electron interaction energy and electron-phonon interaction energy always appear together in Eq.~\eqref{eq:energy_elec}, 
and the electron-electron interaction contribution itself cannot be evaluated from the single-particle Green function.
The difference between Eqs.~\eqref{eq:energy_elec} and \eqref{eq:energy_phon}
needs to be evaluated to obtain $\la \mathscr H_{\rm int} \ra$, and hence we must calculate the phonon self-energy.

\section{Analysis of the normal Fermi liquid}

Before we study the numerical solution of the Eliashberg equations, 
we consider the normal-state properties based on the lowest-order perturbation theory.
The physics in this section is basically the same as previously discussed for the usual electron-phonon coupling \cite{AGD_book}, and the local nature of the phonons in our paper makes it easier to understand the underlying mechanism. 
We first consider the charge-orbital moment correlation function [Eq.~\eqref{eq:def_orb_corr}] given by
\begin{align}
\chi(\imu \nu_m)  &=  4T \sum_n G(\imu\omega_n) G(\imu\omega_n + \imu \nu_m),
\end{align}
where $G$ is the local Green function and the normal state is assumed so that the anomalous Green function is zero ($F=0$).
Here we take the zeroth-order contribution for the Green function:
\begin{align}
    G_0(\imu\omega_n) &= \int \diff \ep  \  \frac{\rho(\ep)}{\imu\omega_n - \ep}.
\end{align}
Using a semi-circular bare density of states, $\rho(\ep) = \frac{2}{\pi D^2} \sqrt{D^2-\ep^2}$, we obtain
\begin{align}
\chi(\imu \nu_m) &= - \frac{16}{3}\rho(0)
+ 4\pi \rho(0)^2 |\nu_m|
\label{eq:analytic_chi}
\end{align}
at low energies and at low temperatures (see Appendix C for the derivation).
The first term corresponds to a static local susceptibility ${\rm Re\,}\chi(0)\propto \rho(0)$.
The second term gives the Korringa relation $\rm Im\, \chi(\omega+\imu 0^+) / \omega \propto \rho(0)^2$ for a normal metal, which originates from a dynamical local correlation function.

This expression immediately yields the phonon self-energy according to Eq.~\eqref{eq:phon_self_def}.
The phonon Green function is given by
\begin{align}
    D_\eta(\imu\nu_m) &= \frac{2\omega_\eta}{(\imu\nu_m)^2 - \tilde \omega_\eta^2 - 2 \gm_\eta |\nu_m|},
    \label{eq:FL_phonon}
\end{align}
where the shifted frequency $\tilde \omega_\eta$ and the damping $\gm_\eta$ for the phonons are given by
\begin{align}
    \tilde \omega_\eta &= \omega_\eta \sqrt{ 1- \frac{32\rho(0)g_\eta^2}{3\omega_\eta} }
    \simeq \omega_\eta - \frac{16}{3}\rho(0)g_\eta^2
    \label{eq:analytic_tilde_omega},
    \\
    \gm_\eta &= 4\pi \rho(0)^2 g_\eta^2 \omega_\eta .
    \label{eq:analytic_gamma}
\end{align}
The decrease in the frequency implies a softening of the phonons caused by the coupling to the electrons.
Indeed, the denominator of Eq.~\eqref{eq:FL_phonon} shows a damped harmonic oscillator with harmonic potential $\frac 1 2 \tilde \omega_\eta^2 x^2$ ($x$ is a coordinate), whose curvature is smaller than the original potential $\frac 1 2 \omega_\eta^2 x^2$
($\omega_\eta > \tilde \omega_\eta$).

We also consider the effect of the phonon self-energy on the effective interaction given in Eq.~\eqref{eq:def_eff_int}.
It evaluates in the low frequency limit to
\begin{align}
    U_{\rm eff}(\imu \nu_m = 0) &= \sum_\eta \left[
    \frac{4}{3} I_\eta -\Big( \frac{\omega_\eta}{\tilde \omega_\eta}\Big)^2
      \lambda_\eta \right] ,
      \label{eq:static_int1}
\end{align}
where we have defined the bare attractive interaction $\lambda_\eta = \frac{4g_\eta^2}{3\omega_\eta}$ 
(not to be confused with the Gell-Mann matrices $\hat \lambda^{\eta}$ in Eq.~\eqref{eq:def_Gell-Mann}).
The effective interaction can be rewritten in terms of the original interaction parameters as
\begin{eqnarray}
    U_{\rm eff}(\imu \nu_m = 0) &=& U 
    -\Big( \frac{\omega_0}{\tilde \omega_0}\Big)^2
      \lambda_0
    \nonumber \\
&&
    +2J
      -2\Big( \frac{\omega_3}{\tilde \omega_3}\Big)^2
      \lambda_3
      -3\Big( \frac{\omega_1}{\tilde \omega_1}\Big)^2
      \lambda_1 , \hspace{3mm}
      \label{eq:static_int2}
\end{eqnarray}
where we have used the cubic symmetry, i.e., the equivalence of $\eta=3,8$ ($E_g$) and that of $\eta=1,6,4$ ($T_{2g}$).
The first and second lines of Eq.~\eqref{eq:static_int2} imply the contributions from the isotropic ($\eta=0$) and anisotropic parts ($\eta\neq 0$), respectively.
The repulsive Coulomb interaction implies that the contribution from $U$ ($>0$) is not favorable for Cooper pairs, and $J$ ($>0$) is energetically unfavorable for spin-singlet pairing.
In the electron-phonon interaction parts,
since there is the relation $\omega_\eta > \tilde \omega_\eta$ due to the softening of the phonons, the attractive interaction is expected to be enhanced by the phonon self-energy.
We will further examine this point with numerical calculations in the next section (see Sec.~V D).

It can be recognized that the Jahn-Teller phonon contributions $\lambda_{1,3,4,6,8}$ are important as they give five times larger contributions in total in the present system, compared to the isotropic component $\lambda_0$ ($\sim \lambda_1$, see Appendix A).
In contrast, for the electron-electron interaction, the anisotropic parts $I_{1,3,4,6,8}$ are of the order of the Hund's coupling $J$, which is small compared to the isotropic component $I_0 \sim U$, especially in fullerides with extended molecular orbitals.
Hence, compared with the isotropic case which has only the $\eta=0$ component in the first line of Eq.~\eqref{eq:static_int2}, the effective attractive interaction is much enhanced in the system with Jahn-Teller phonons. This gives a simple answer why the fulleride superconductors with multiple degrees of freedom have a high transition temperature.

Once the phonon propagator has been obtained, the self-energy for the electrons can be also evaluated as
\begin{align}
    \Sigma(\imu\omega_n) 
    =& - \imu \rho(0)\omega_n \sum_\eta
    \frac{\omega_\eta^2 \lambda_\eta}{\tilde \omega_\eta^2}
    \nonumber \\
    &  
    - \imu \rho(0)(\pi^2 T^2  - \omega_n^2) 
    \, {\rm \sgn}\omega_n
    \sum_\eta \frac{ \omega_\eta^2 \gm_\eta \lambda_\eta }{\tilde \omega_\eta^4},
    \label{eq:analytic_sigma}
\end{align} 
where we have assumed the magnitude relations $T, |\omega_n|\ll \omega_\eta \ll D$
(see Appendix C for the derivation).
This expression determines the properties of the Fermi liquid such as the mass renormalization (first term) and quasi-particle damping (second term) near zero temperature.
While the above expressions are based on several assumptions such as $\lambda_\eta, \omega_\eta \ll D$, they fit the numerical results well at small electron-phonon coupling, as shown in the next section.
On the other hand, the electron-electron interaction only gives a chemical potential shift in the normal state and does not alter the physical properties.
We will reconsider this point in Sec.~VI and take fluctuations from $I_\eta$ into account.

\section{Numerical solution of the Eliashberg equation}

\subsection{Parameters}

Throughout this section, we take the energy unit $2D=1$, where $D$ is the half band width of the semi-circular density of states 
($D$ will be varied in Sec.~VI).
We consider the three half-filled orbitals, as realized in the alkali-doped fullerides, and choose $\lambda_\eta = \lambda_0$.
This condition with no $\eta$-dependence approximately corresponds to the realistic situation for fullerene molecules (see Appendix A for more details).
The Hund's coupling is fixed as $J/U\simeq 0.03$ \cite{Nomura16}, and we choose $\omega_\eta = \omega_0 = 0.15$ in the following.
The tuning parameters are then the Coulomb interaction $U$, the electron-phonon coupling $\lambda_0$, and the temperature $T$.
Since the goal of this paper is a qualitative understanding of  superconductivity in fullerides, we systematically vary these tuning parameters.

\subsection{Normal state}

\begin{figure}[tb]
    \centering
    \includegraphics[width = 80mm]{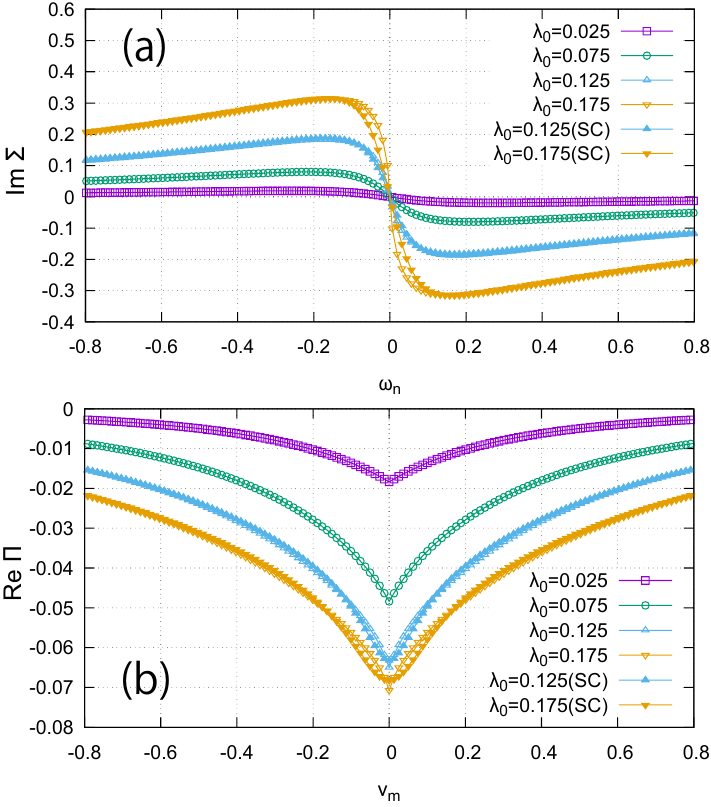}
    \caption{
    (a) Electron self-energy and (b) phonon self-energy evaluated in the normal and superconducting states. The parameters are chosen as $T=0.002$ and $U=2$.
    The symbol `SC' indicates a solution in the superconducting state.
    }
    \label{fig:self-energy}
\end{figure}

We begin with the discussion of the frequency dependence of the self-energies.
We take 
$T=0.002$, and change the electron-phonon coupling $\lambda_0$ ($=\lambda_\eta$ for all $\eta$).
The system becomes  superconducting for $\lambda_0 = 0.125$ and $0.175$ if we allow for a nonzero pair potential.
Here we concentrate on the normal state, while the pairing state will be discussed in the next subsection.
In the normal phase, the Coulomb interaction $U$ just contributes to the chemical potential shift and may be neglected, while it affects the pair potential.

Figure~\ref{fig:self-energy}(a) shows the imaginary part of the electron normal self-energy on the imaginary-frequency axis.
The real part is zero at half-filling.
With increasing  electron-phonon coupling, the magnitude of the self-energy is enhanced.
The linear slope at low frequencies also becomes steeper for larger $\lambda_0$, indicating that the renormalization of the quasiparticles is stronger.

We also show the phonon self-energy in Fig.~\ref{fig:self-energy}(b).
The static component at $\nu_m = 0$ shows the energy shift caused by the electron-phonon coupling.
In the normal state, the low-energy limit has a linear functional form, although the whole function is even with respect to frequency.
This non-analytic behavior corresponds to the damping which originates from the second term in Eq.~\eqref{eq:analytic_chi}.

\begin{figure}[tb]
    \centering
    \includegraphics[width = 75mm]{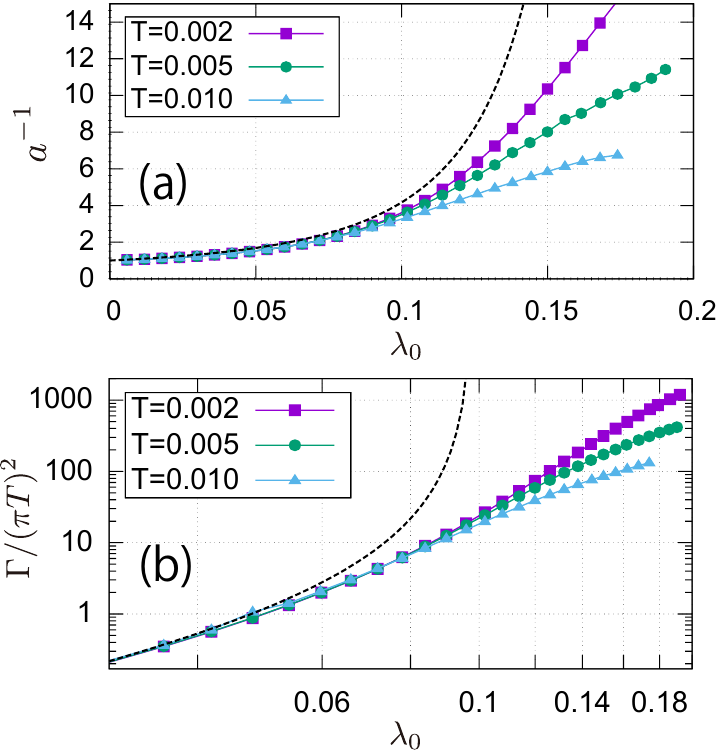}
    \caption{
    Electron-phonon interaction parameter dependence of (a) the inverse of the renormalization factor and (b) the damping factor divided by the square of temperature,  defined in Eqs.~\eqref{eq:elec_renorm} and \eqref{eq:elec_damp}, evaluated in the normal state.
    The dotted lines indicate the analytic form given in Eq.~\eqref{eq:analytic_sigma}.
}
    \label{fig:temp_dep_electron}
\end{figure}

Next, we show the characteristic coefficients extracted from the self-energies.
The renormalization factor $a$ of the electrons is defined by
\begin{align}
    a &= \left( 1 - \lim_{\omega_n\to 0} \frac{\partial \imag \Sigma(\imu\omega_n)}{\partial \omega_n} \right)^{-1} \leq 1
    \label{eq:elec_renorm}
\end{align}
and the quasiparticle damping $\Gamma$ by
\begin{align}
    \Gamma &= - \lim_{\omega_n\to +0} \imag \Sigma(\imu\omega_n) \geq 0 .
    \label{eq:elec_damp}
\end{align}
The actual extrapolation to zero frequency is performed numerically at each temperature.
Hence the extrapolated values are meaningful only at low temperatures.

Figure~\ref{fig:temp_dep_electron} shows the inverse of the renormalization factor in (a) and the damping divided by the square of temperature in (b) as a function of $\lambda_0$ at several temperatures.
In panel (b), we plot the data on a logarithmic scale since the damping covers a wide range of values.
Both quantities are increasing functions of the electron-phonon coupling.
At small couplings,  $\lambda_0 \lesssim 0.1$, the renormalization factor is temperature-independent as shown in Fig.~\ref{fig:temp_dep_electron}(a) at low temperatures, indicating the formation of a Fermi liquid.
The damping also shows a Fermi-liquid behavior proportional to the square of temperature.
In this weak-coupling regime, the behaviors are approximately captured by the analytic formula in Eq.~\eqref{eq:analytic_sigma}.

On the other hand, for $\lambda_0 > 0.1$, the value of $a^{-1}$ 
shows a significant temperature dependence 
even at low temperatures.
This implies a non-Fermi liquid behavior at least in the temperature regime shown in the figure.
This anomaly is also reflected in other physical quantities, such as the specific heat, as will be shown later.
While the calculations have been performed by assuming a normal state (i.e., $\Delta(\imu\omega_n) = 0$), 
these non-Fermi liquid solutions
would be replaced by a superconducting phase if we allow for a nonzero $\Delta$. 
The anomaly at low temperatures in the normal state may however still be observed under a magnetic field which suppresses the superconducting state.

Actually, the non-Fermi liquid behavior in the normal state seems to be unavoidable for a strong-coupling superconductor: it is closely related to the entropy balance seen in the specific heat.
We will consider this aspect later in Sec.~V C.
We note that, in the context of the present paper, the terminology `strong coupling' means a strong retardation effect from the electron-phonon coupling \cite{Parks_book}, which is different from the higher-order electronic correlation effects.

\begin{figure}[tb]
    \centering
    \includegraphics[width = 75mm]{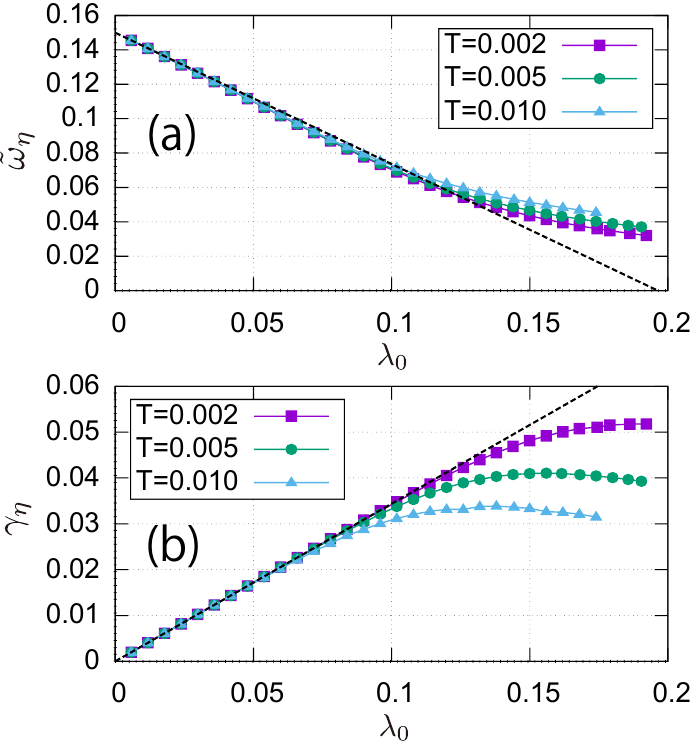}
    \caption{
    Electron-phonon interaction parameter dependence of (a) the shifted phonon energy determined by Eq.~\eqref{eq:phon_shift} and (b) the phonon damping determined by Eq.~\eqref{eq:phon_damp}, evaluated in the normal state.
    The dotted lines show the analytic functional forms of Eqs.~\eqref{eq:analytic_tilde_omega} and \eqref{eq:analytic_gamma}.
}
    \label{fig:temp_dep_phonons}
\end{figure}

In the case of the phonons, the shifted quasiparticle energy $\tilde \omega_\eta$ and damping $\gm_\eta$ are extracted from the low-energy properties of the self-energy:
\begin{align}
    \tilde \omega_\eta^2 &= \omega_\eta ^2 + 2\omega_\eta \lim_{\nu_m \to +0}  {\rm Re\,} \Pi(\imu \nu_m) ,
    \label{eq:phon_shift}
    \\
    \gm_\eta &= \omega_\eta \lim_{\nu_m \to +0} \frac{\partial {\rm Re\,} \Pi(\imu \nu_m)}{\partial \nu_m} .
    \label{eq:phon_damp}
\end{align}
These parameters are shown in Fig.~\ref{fig:temp_dep_phonons}.
The phonon energy is shifted to a smaller value as shown in panel (a), which indicates a softening of the phonons by the coupling to the electrons.
As for the damping shown in panel (b), it increases linearly in the Fermi-liquid regime at small couplings, while the behavior changes at large $\lambda_0$.
The temperature dependence is still present at large coupling, reflecting the non-Fermi liquid behavior, and is stronger for the damping in panel (b), compared to the shifted energy in panel (a).

\subsection{Superconducting state}

\begin{figure}[tb]
    \centering
    \includegraphics[width = 85mm]{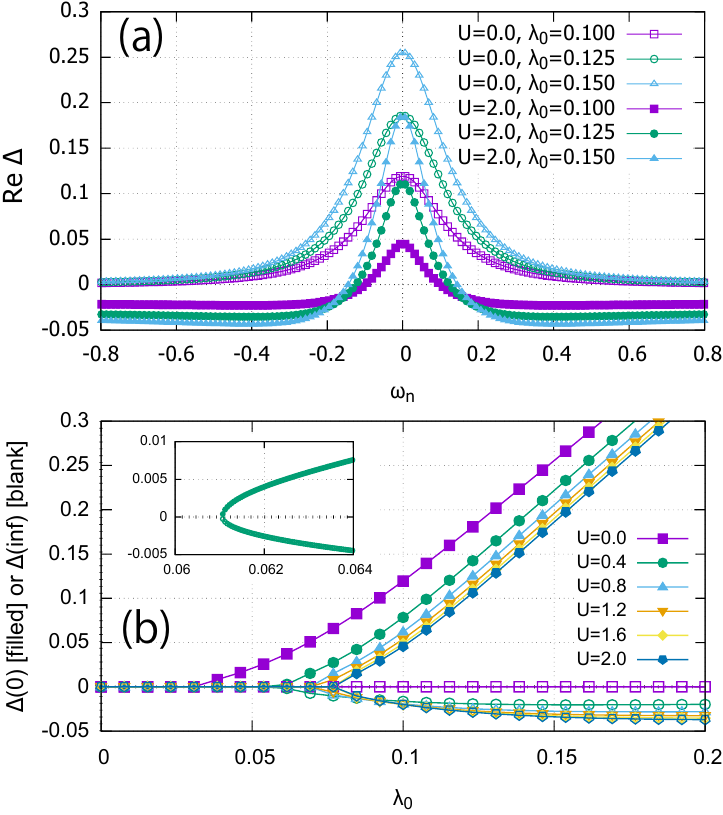}
    \caption{
    (a) Anomalous self-energy for the electrons.
    (b) Electron-phonon interaction parameter dependence of the pair potential in the  low-frequency (filled symbols) and high-frequency (open symbols) limits.
    The temperature is chosen as $T=0.002$.
    The inset shows a zoom of the region near the transition point for $U=0.4$.
    }
    \label{fig:anomalous_self}
\end{figure}

In this subsection, we discuss the frequency dependence of the anomalous self-energy shown in Fig.~\ref{fig:anomalous_self}(a). It has the dimension of energy and is sometimes referred to as pair potential or gap function.
A larger $\lambda_0$ results in a larger magnitude, reflecting the stronger effective attraction among the electrons.
For $U=0$, the anomalous self-energy is largest at zero frequency, and becomes zero in the high-frequency limit.
This behavior is characteristic of the retardation effect caused by the electron-phonon coupling and is in contrast with the usual BCS theory, which leads to a frequency-independent gap function.
For $U>0$, the magnitude at low frequencies is decreased by the repulsive Coulomb interaction.
It is also notable that the signs in the low-frequency limit and high-frequency limit are opposite, which indicates a change from attractive to repulsive interactions 
at a certain frequency.

The effect of the pair potential is also reflected in the normal self-energy as shown in Fig.~\ref{fig:self-energy}.
Although the superconducting state is realized for $\lambda_0 = 0.125$ and $0.175$, its effect on Im$\Sigma$ is not prominent as shown in Fig.~\ref{fig:self-energy}(a).
In the phonon self-energy, on the other hand, a qualitatively different feature can be seen for the superconducting case with $\lambda_0=0.125,0.175$: the linear behavior at low frequencies changes to a quadratic one as shown in Fig.~\ref{fig:self-energy}(b).
This reflects the energy gap formation in the electronic state, which removes the damping in the low-energy limit.
However, the difference between normal and superconducting states is small, which agrees with the expectation from Ref.~\cite{Bardeen64}.

The anomalous self-energy in Fig.~\ref{fig:anomalous_self}(a) is characterized by the two values at low frequency [$\Delta(\imu \omega_n=0)$] and at high frequency [$\Delta(\imu \omega_n=\infty)$].
The $\lambda_0$-dependence of these quantities is shown in Fig.~\ref{fig:anomalous_self}(b) at several values of $U$.
The magnitude of $\Delta(0)$ continues to increase with increasing electron-phonon coupling, while $\Delta(\infty)$, which originates from the Coulomb interaction, does not.
Although the critical behavior near the transition point is of the square-root mean-field-type, as shown in the inset of panel (b), such a dependence can only be seen in a narrow parameter regime.

\subsection{Specific heat}

\begin{figure}[t]
    \centering
    \includegraphics[width = 85mm]{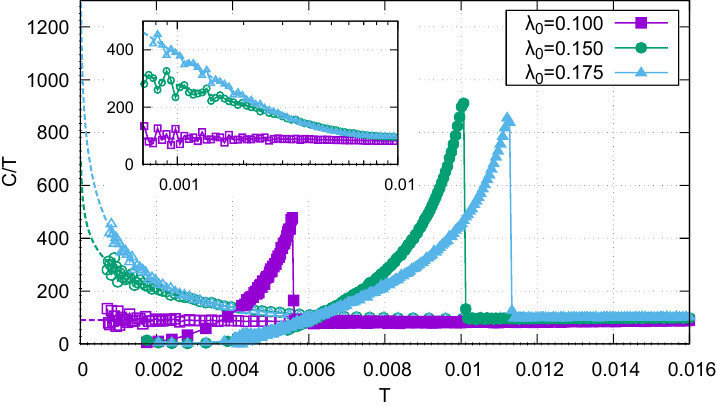}
    \caption{
Temperature dependence of the specific heat divided by the temperature.
The interaction parameter is taken as $U=2$.
The inset shows the data for the normal phase on a logarithmic temperature scale.
The filled symbols show the results for the superconducting state, while the open symbols are for the normal state.
The dotted lines are a guide to the logarithmic temperature dependence.
    }
    \label{fig:tdep_spec_heat}
\end{figure}

We now consider the specific heat which is evaluated as the temperature derivative of the internal energy: $C = \frac{\partial \la \mathscr H \ra}{\partial T} = \frac{\partial}{\partial T} \big( \la \mathscr H_{\rm e0}\ra+\la\mathscr H_{\rm int}\ra+\la\mathscr H_{\rm p}\ra + \la\mathscr H_{\rm ep}\ra \big)$.
Solving the self-consistent equations, we obtain the self-energies for the electrons and phonons.
Then, the interaction energies can be calculated by Eqs.~\eqref{eq:energy_elec} and \eqref{eq:energy_phon}.
The kinetic and potential energies can also be evaluated from the local Green functions.
Thus the internal energy is calculated from the single-particle Green functions and self-energies.
We note that both the electron and phonon self-energies are needed in our algorithm.

Figure~\ref{fig:tdep_spec_heat} shows the temperature dependence of the specific heat at $U=2$.
We first consider the normal state.
The results are shown by open symbols, and smoothly change down to the lowest temperature.
The fluctuating behavior originates from a numerical error associated with the numerical temperature derivative, although the internal energy itself is an almost smooth function.
On the high-temperature side of the figure, $C/T$ behaves as nearly temperature-independent.
In the weak coupling case, e.g. $\lambda_0=0.1$, the specific heat coefficient is temperature-independent with the value $C/T\simeq 100$ down to the lowest temperature.
In this case, the renormalization factor is roughly estimated from Fig.~\ref{fig:temp_dep_electron}(a) as $a^{-1}\sim 4$.
Since the non-interacting limit of the specific heat coefficient evaluates to $C/T = 2\pi^2 \rho(0) \simeq 25$, the expected $C/T$ value from the Fermi liquid theory is $4\times 25=100$, consistent with the specific heat calculation.
For a stronger coupling, however, the situation changes:
the specific heat shows a logarithmic increase at low temperatures, as shown in the inset, which clearly deviates from the Fermi liquid behavior.

If we consider the superconducting state, the logarithmic region is replaced by the superconducting region (see filled symbols).
Although the system above the transition temperature shows a nearly temperature-independent $C/T$, 
the renormalization factor plotted in  Fig.~\ref{fig:temp_dep_electron}(a) shows a strong temperature dependence.
Hence, the transition at $T=T_c$ does not occur from a normal Fermi liquid.

Next, we focus on the entropy balance: both in the normal (n) and superconducting (s) state, the entropy must be the same at the transition temperature:
\begin{align}
    S_{\rm n}(T_c) - S_{\rm s}(T_c) &= \int_0^{T_{\rm c}} \diff T \frac{C_{\rm n}(T) - C_{\rm s}(T)}{T} = 0 .
\end{align}
This quantity may be  graphically estimated as the closed area shown in Fig.~\ref{fig:tdep_spec_heat}.
For a small coupling such as $\lambda_0 = 0.1$, the entropy balance is nearly satisfied with a constant (Fermi-liquid) $C/T$ for the normal state.
On the other hand, for larger couplings, the balance would not be satisfied if $C_{\rm n}/T$ were temperature-independent below $T_c$.
This is because the specific heat jump is larger in the strong-coupling case.
Hence the normal state specific heat must be enhanced at low temperatures to balance the result of the superconducting state.
Thus, the logarithmic divergence of the specific-heat coefficient appears to be inevitable for 
a superconductor with strong retardation effects.

\begin{figure}[tb]
    \centering
    \includegraphics[width = 85mm]{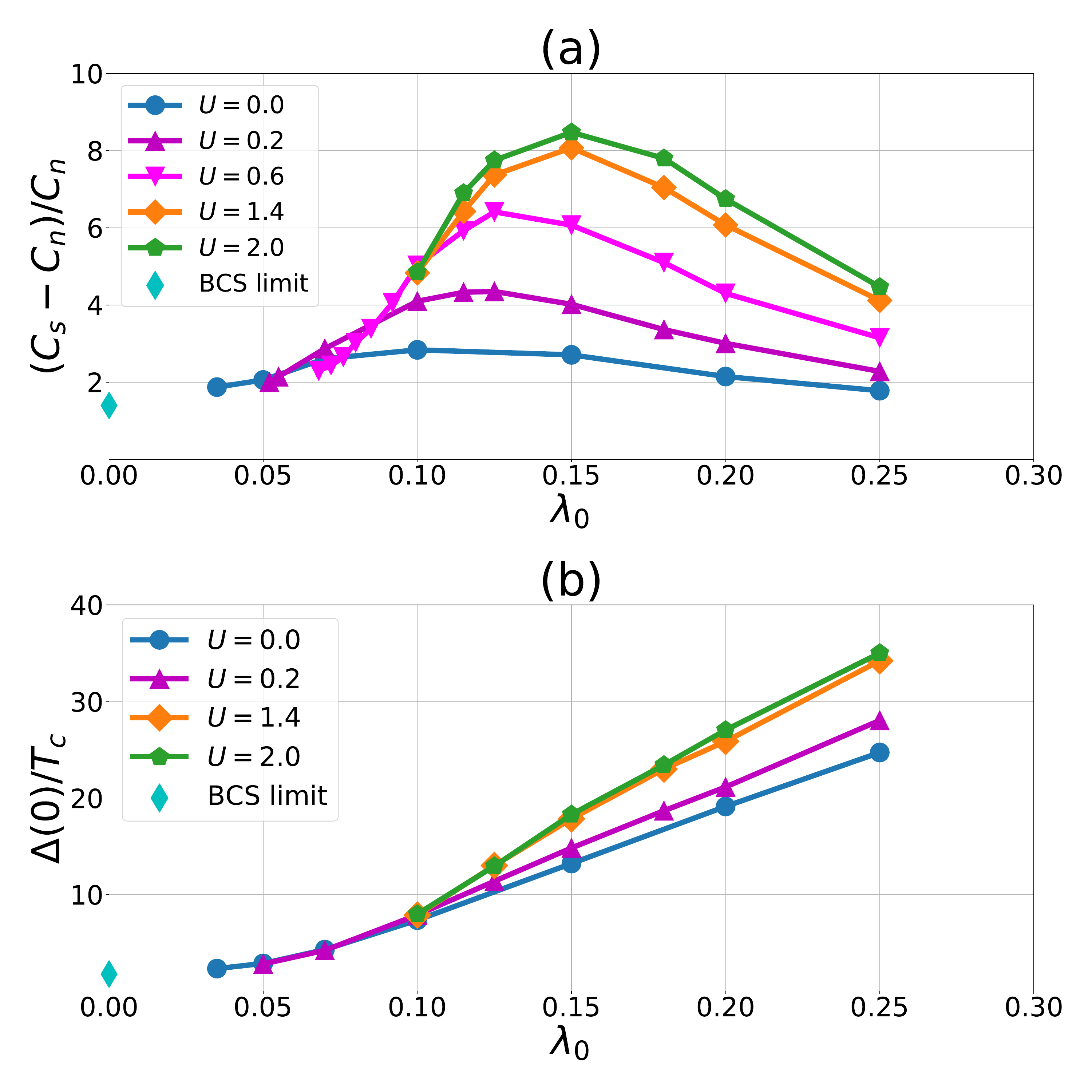}
    \caption{
    (a) Specific heat jump as a function of $\lambda_0$ evaluated at $T=T_c$.
    The value is normalized by the specific heat in the normal state at the transition temperature.
    (b) Ratio between the gap function in the low-frequency limit and the transition temperature.
    }
    \label{fig:specific_heat}
\end{figure}

For a strong-coupling superconductor, the specific-heat jump at the transition temperature is known to be enhanced, compared to the BCS limit \cite{Parks_book}.
Here we also confirm this behavior for the Jahn-Teller-Hubbard model: Figure~~\ref{fig:specific_heat}(a) shows the specific heat jump  
$C_{\rm s}(T_c) - C_{\rm n}(T_c)$ normalized by $C_{\rm n} (T_c)$.
We also add a point for the BCS limit evaluated by the attractive model with $U=-0.2$.
The value in the weak-coupling limit approaches this BCS value.
As a related quantity, we also show the ratio between $\Delta(0)$ and $T_c$ in Fig.~\ref{fig:specific_heat}(b), which is also enhanced compared to the universal value in the BCS theory.
The ratio $\Delta(0)/T_c$ is a simple increasing function of $\lambda_0$, while the specific-heat-jump decreases on the stronger coupling side.

\subsection{Phase diagram}

\begin{figure}[tb]
    \centering
    \includegraphics[width =85mm]{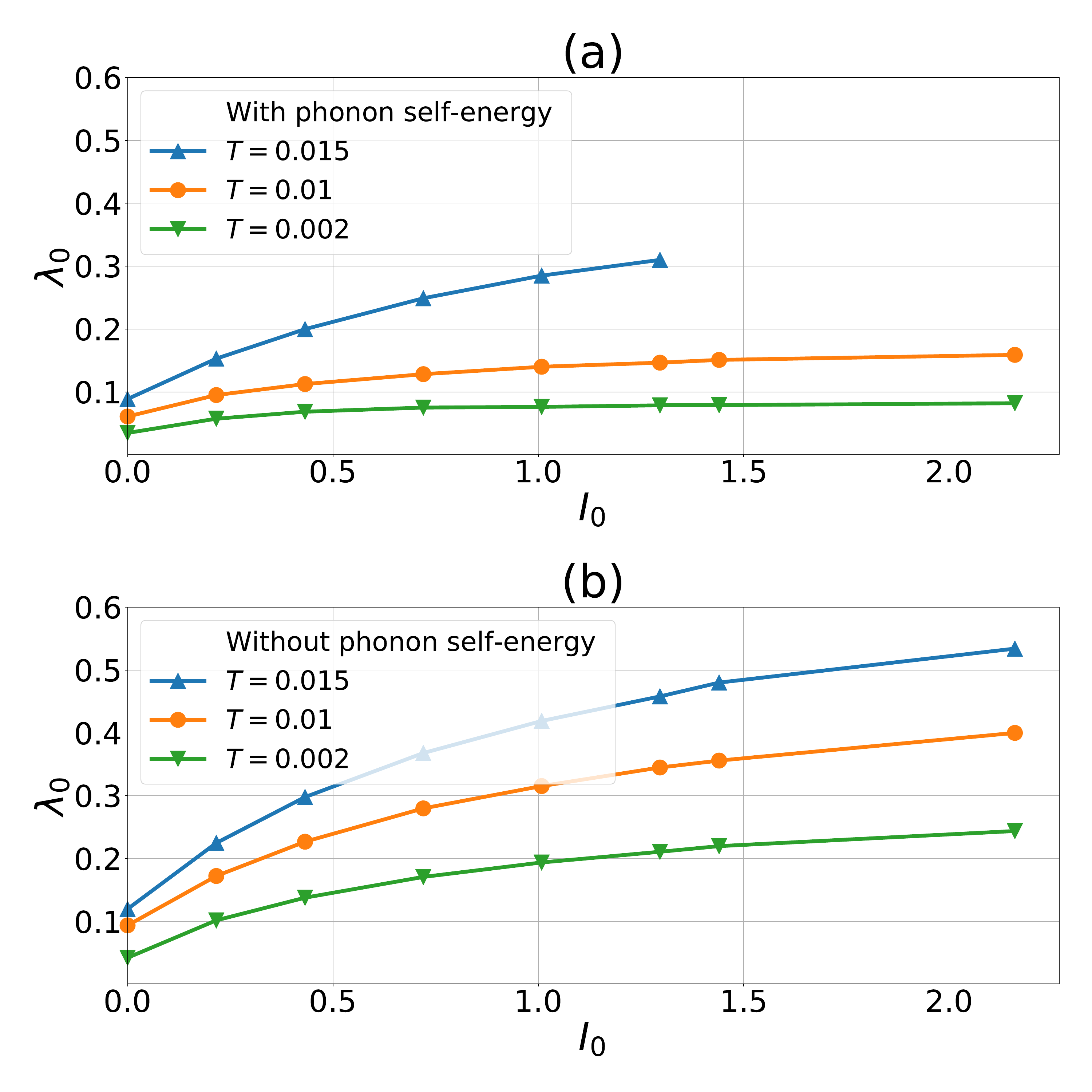}
    \caption{
    Phase diagram in the plane of $I_0$ and $\lambda_0$.
    The phonon self-energy is considered in the top panel (a), while it is neglected in the bottom panel (b).
    We take $\omega_\eta = 0.15$ and $\lambda_\eta = \lambda_0$.
}
    \label{fig:phase1}
\end{figure}

Figure \ref{fig:phase1}(a) shows the $I_0$-$\lambda_0$ phase diagram in the plane of the interaction parameters.
The region above the phase boundary becomes superconducting due to the attractive interaction caused by the electron-phonon coupling $\lambda_0$.
This phase boundary shifts  downward at lower temperatures to enlarge the superconducting regime.

It is interesting to compare these results with those for zero phonon self-energy ($\Pi=0$). 
We note that, with the above approximation, Eq.~\eqref{eq:energy_elec} is not identical to Eq.~\eqref{eq:energy_phon}  
without 
electron-electron interactions.
Whereas the self-consistent treatment of the internal energy is lost, the solution can still be obtained by the iterative method using the Eliashberg equation.
The results are shown in the panel (b) of Fig.~\ref{fig:phase1} and demonstrate the importance of the phonon self-energy.
We need larger values of the electron-phonon coupling to induce the superconducting state, compared with the case with phonon self-energy.
Hence, the phonon self-energy helps the superconducting state by softening the phonon frequency, as discussed in connection with Eq.~\eqref{eq:static_int2}.

\section{Phonon-orbiton coupling}

We now proceed one step further in order to improve the accuracy of the theory.
Utilizing the fact that the electron-electron interaction \eqref{eq:ham_ep} and electron-phonon interaction \eqref{eq:int_rewrite} are written in terms of the same physical quantity $T_{i\eta}$ (charge-orbital moments), we try to incorporate the fluctuations from the Coulomb interaction.
First of all, we define the momentum of the phonons 
\begin{align}
p_{i\eta} = (a_{i\eta}-a_{i\eta}^\dg) / \imu ,
\end{align}
which satisfies the canonical commutation relation $[\phi_{i\eta}, p_{i\eta}] = 2\imu$.
With this, the action can be rewritten as
\begin{align}
     &S[\phi,p,c] = S_{e0}[c] + 
     \sum_{i \eta}\int \diff \tau 
     \Big[
    - \frac \imu 2 p_{i\eta} \partial_\tau \phi_{i\eta}
    \nonumber \\
     &\hspace{10mm}
    +\frac {\omega_\eta}{4} (\phi_{i\eta}^2 + p_{i\eta}^2)
    + g_\eta \phi_{i\eta} T_{i\eta}
    + I_\eta T_{i\eta}^2
    \Big] ,
\end{align}
where $I_\eta > 0$ (the case with $I_\eta <0$ can be considered separately).
Given the above form of the interaction terms, it is tempting to describe the electron-electron interaction and electron-phonon interaction in a unified way.
The Stratonovich-Hubbard transformation for the Coulomb interaction term is unique for this representation, and the integral with respect to the $p$-field gives
\begin{align}
    S'[\phi,Q,c] &= S_{e0}[c] +  \sum_{i \eta}\int \diff \tau  \Big[
    \frac {\omega_\eta}{4} \phi_{i\eta}^2 
    + \frac{1}{4\omega_\eta} (\partial_\tau \phi_{i\eta})^2
    \nonumber \\
    &\ \ 
    + \frac{I_\eta}{2}  Q_{i\eta}^2
    + (g_\eta \phi_{i\eta} + \sqrt{2}\,\imu I_\eta Q_{i\eta}  ) T_{i\eta}
    \Big] .
    \label{eq:action_phi_Q_c}
\end{align}
The newly introduced local variable $Q_{i\eta}$ is similar to the plasmon in the electron gas problem \cite{Nagaosa_book}.
This bosonic variable is associated with the electronic orbital moment, and we call it ``local orbiton.''
(Since the $Q_{i,\eta=0}$ component originates from the local charge, we may call it ``local plasmon.'')
In the spirit of DMFT, we include the local self-energy contribution of this term.
From Eq.~\eqref{eq:action_phi_Q_c}, we recognize that the electron operator is coupled to a phonon-orbiton field $\tilde \phi$ 
defined as 
\begin{align}
    \tilde g_\eta \tilde \phi_{i\eta}(\tau) \equiv g_\eta \phi_{i\eta}(\tau) + \sqrt{2}\,\imu I_\eta Q_{i\eta}(\tau) ,
    \label{eq:modified_coupling}
\end{align}
where $\tilde g$ may be regarded as a renormalized coupling constant.
Then the formulation in the last sections can be recycled, by replacing $g\to \tilde g$ and $\phi \to \tilde \phi$.

The electronic part can be integrated with the use of second-order 
perturbation theory.
In the frequency domain, the effective action for the bosons is given by
\begin{align}
    &S''[\phi,Q] = \sum_{i m\eta}
\begin{pmatrix}
    \phi^*_{i\eta}(\imu\nu_m) & Q^*_{i\eta}(\imu\nu_m)
\end{pmatrix}
\nonumber \\
&\ \ \times
\frac{1}{2}
\begin{pmatrix}
    \frac{\omega_\eta}{2} - \frac{1}{2\omega_\eta}(\imu\nu_m)^2 + g_\eta^2 \chi(\imu\nu_m) 
    & \sqrt 2 \, \imu g_\eta I_\eta \chi  (\imu\nu_m)
    \\[1mm]
    \sqrt 2 \, \imu g_\eta I_\eta \chi (\imu\nu_m)
    & I_\eta - 2I_\eta^2 \chi(\imu\nu_m)
\end{pmatrix}
\nonumber \\
&\ \  \times 
\begin{pmatrix}
    \phi_{i\eta}(\imu\nu_m) \\ Q_{i\eta}(\imu\nu_m)
\end{pmatrix}
\\
&= \sum_{i m \eta} 
\vec \phi_{i\eta} ^\dg(\imu\nu_m)
\big[  - \tfrac 1 2 \hat D_\eta^{-1} (\imu\nu_m)  \big]
\vec \phi_{i\eta}(\imu\nu_m)
\end{align}
where 
$\hat D_\eta = 
\begin{pmatrix} 
D_{\eta, \phi\phi} & D_{\eta,\phi Q}\\ 
D_{\eta,Q\phi} & D_{\eta,QQ}
\end{pmatrix} $
with $D_{\eta,AB}(\tau) = - \la \mathcal T A_{i\eta}(\tau) B_{i\eta}(0) \ra$
 is the bosonic Green function matrix of the coupled phonon-orbiton system.
The orbital correlation function $\chi$
has already been defined in Eq.~\eqref{eq:def_orb_corr}.
The off-diagonal components show that
the phonon and orbiton are coupled locally through the coupling to the electrons.
If we set $I_\eta$ to zero, we reproduce the results in Sec.~III.

We can also calculate the electronic self-energy from Eqs.~\eqref{eq:simplified_sigma} and \eqref{eq:simplified_delta}.
Considering  Eqs.~\eqref{eq:modified_effective_interaction} and \eqref{eq:modified_coupling}, the effective interaction may be written as
\begin{align}
\tilde {\mathscr U}_{\eta}(\tau) &= \frac 1 2  \tilde g_\eta^2 D_{\eta, \tilde \phi \tilde \phi}(\tau)
\\
&=
\frac 1 2 g_\eta^2 D_{\eta,\phi\phi}(\tau)
+ \sqrt 2 \, \imu g_\eta I_\eta D_{\eta,\phi Q}(\tau)
- I_\eta^2 D_{\eta,QQ}(\tau)
.
\label{eq:eff_int_orbiton}
\end{align}
Note that this expression is based on second-order perturbation theory with respect to the coupling between the electrons and bosons.

With the coupling to orbitons, a dynamical correction to the electron self-energy appears even without the coupling to phonons.
In order to recognize the effect which is newly included here, let us consider the case with $g_\eta = 0$.
The effective interaction for the channel $\eta$ can then be written as
\begin{align}
\tilde {\mathscr U}_\eta (\imu \nu_m) 
= \frac{I_\eta}{1 - 2 I_\eta \chi(\imu \nu_m)},
\label{eq:eff_int_Coul_fluct}
\end{align}
where $\tilde {\mathscr U}_\eta (\infty) =I_\eta$ corresponds to a bare interaction ($\chi < 0$).
Since the relation $0< \tilde {\mathscr U}_\eta(\imu\nu_m) < I_\eta$ holds, the effective interaction $\tilde {\mathscr U}(\imu\nu_m)$ represents a locally 
screened Coulomb repulsive interaction.
The different $\eta$-channels are not mixed in this expression.
It follows from Eq.~\eqref{eq:eff_int_Coul_fluct} that the screening factor
$(1-2I_\eta \chi)^{-1}$ becomes smaller if the original interaction $I_\eta$ is bigger.
This 
indicates that the charge part $I_0 \sim U$ is strongly screened, while the effect for the orbital part $I_{1,3,4,6,8}\sim J$ ($\ll U$) is much weaker.

\begin{figure}[tb]
    \centering
    \includegraphics[width = 85mm]{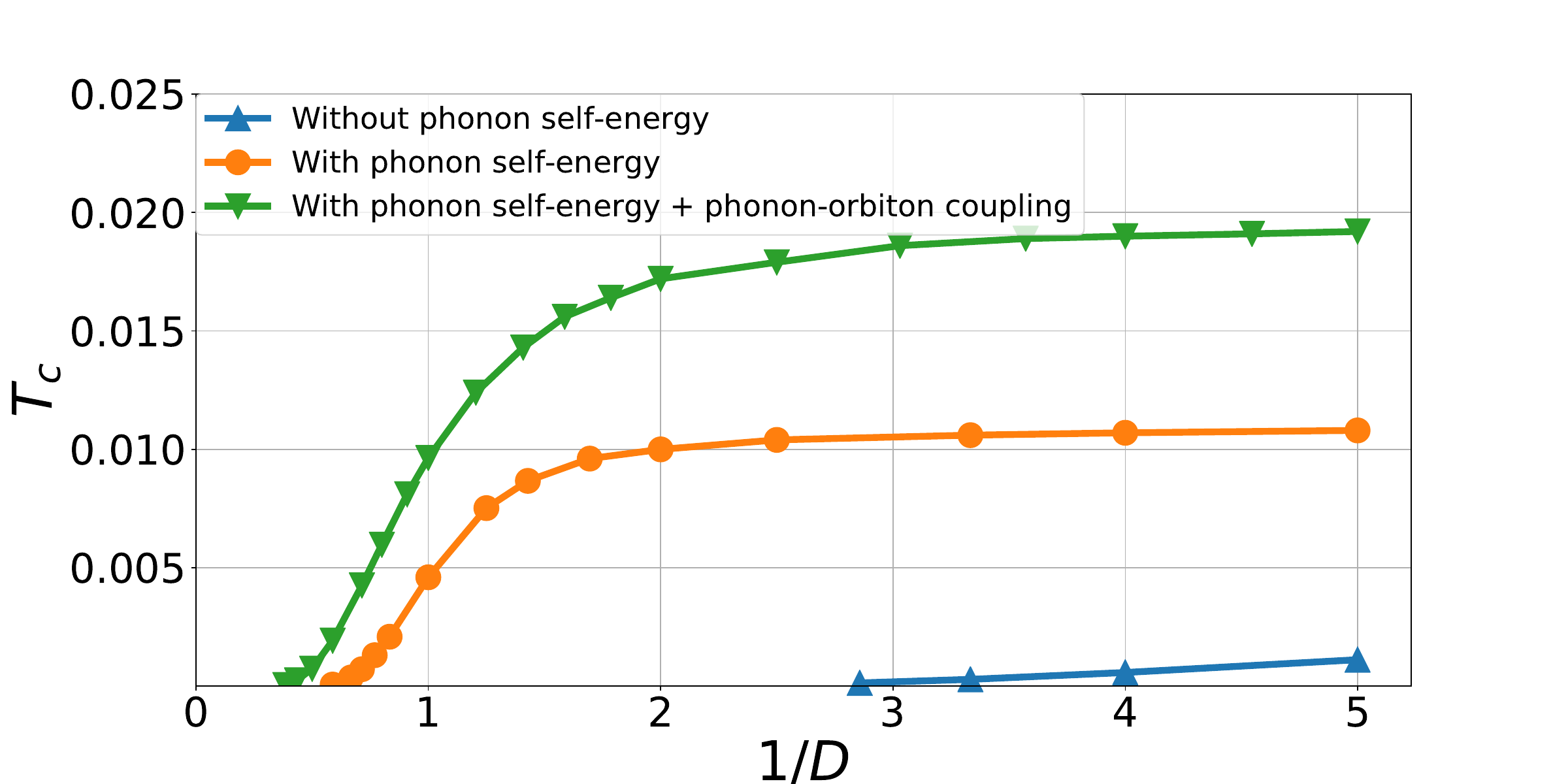}
    \caption{
    Phase diagram in the plane of temperature and inverse
    band-width.
    The parameters are chosen as $U=2$ and $\lambda_0 = 0.15$.
    }
    \label{fig:phase3} 
\end{figure}

Figure \ref{fig:phase3} shows the transition temperature obtained (i) without phonon self-energy, (ii) with phonon self-energy, and (iii) by adding the phonon-orbiton coupling.
The horizontal axis qualitatively corresponds to the 
negative
pressure effect, which diminishes the overlap of the molecular orbitals of the fullerides.
From the figure, we conclude that the transition temperature is enhanced by the phonon self-energy, and is further pushed up by the coupling to orbitons.
The latter occurs because the fluctuation effects from the Coulomb interaction 
effectively reduce the bare repulsive interaction.
Note that this result applies only to the weak-coupling side of the pressure-temperature phase diagram, while strong correlation effects must be taken into account for the other side close to the Mott insulator.

Let us emphasize again that the present approach is based on perturbation theory and the results are not necessarily physical when they are applied to large coupling constants.
For example, the system becomes superconducting even without the coupling to phonons
(at $D=0.5$ and $T=0.002$,  superconductivity appears for $U\gtrsim 8$).
This is because the fluctuation effect included in the above formulation always gives an attractive contribution to the effective interaction. 
Hence, the results are valid only at small Coulomb interactions, while the region with larger couplings requires more sophisticated techniques.
Despite this limitation, the approach formulated in this paper is useful for the physical interpretation of the solutions.

\section{Summary and discussions}

In this paper, we have studied the Jahn-Teller-Hubbard model, where the electrons in multiple orbitals are locally coupled to both isotropic and anisotropic phonons.
We employed the Eliashberg approach and calculated the self-energies for the electrons and phonons.
We discussed the characteristic behaviors of the specific heat and mapped out the phase diagram.
The effect of the coupling to the anisotropic phonons and their self-energy is important to account for the superconducting transition temperature.
The fluctuations from the Coulomb interaction may be incorporated by introducing the orbitons through the Stratonovich-Hubbard transformation of the Coulomb interaction, where the phonons and orbitons can be handled in a unified way because of the charge-orbital moment description of the interactions.

The formulation given in this paper may be understood in analogy to the electron gas model with infinitesimal translational symmetry.
There, the electron-electron interaction can be written as
\begin{align}
    \mathscr H_{\rm int} &= \int \diff \bm q \, I(\bm q) n(\bm q) n(-\bm q)
\end{align}
and the electron-phonon coupling as
\begin{align}
    \mathscr H_{ep} &= \int \diff \bm q \, g({\bm q}) \phi ({\bm q}) n(-\bm q) ,
\end{align}
where $n(\bm q)$ is the Fourier transformation of the electron density, and $\phi(\bm q)$ that of the local dilation \cite{Kittel_book}.
These expressions for the  electron gas model are analogous to the expressions in Eqs.~\eqref{eq:int_rewrite} and \eqref{eq:ham_ep}, if the center of mass momentum $ \bm q$ is replaced by the index $\eta$.
The crucial step to manipulate the electron-electron interaction and electron-phonon coupling is to rewrite the interaction in a form that respects the symmetry, i.e., the multipole representation in Eq.~\eqref{eq:int_rewrite} \cite{Iimura21}.
Thus this representation is useful beyond a simple rewriting of the interaction form, and connects to the concept of momentum in the electron gas model.

We have found a low-temperature logarithmic divergence of the specific heat in the normal state,  which is necessary to assure the entropy balance between the normal and superconducting states with strong retardation effects.
In this context, it is interesting to point out that the recently discovered superconductor UTe$_2$ \cite{Ran19, Aoki19} shows a peculiar behavior of the specific heat: the entropy balance seems not satisfied within the experimentally measured temperature range.
Even though the pairing mechanism should be different, our results suggest that the strong coupling nature requires the normal-state specific-heat coefficient to be enhanced at very low temperatures.

\section*{Acknowledgement}

S.H. would like to thank T. Miki and H. Shinaoka for useful discussions.
This work was supported by JSPS KAKENHI Grants No. JP18H01176, No. JP19H01842, and No. JP21K03459. P.W. acknowledges support from SNSF Grant No.~200021-196966.

\appendix

\section{Electron-phonon coupling in fullerene molecules}

In this appendix, we discuss the electron-phonon system in an isolated C$_{60}$ molecule, which has $3\times 60 - 6$ vibration modes.
The $t_{1u}$ electrons can be coupled with two $A_{g}$ phonons and eight $H_g$ phonons (point group $I_h$) \cite{Chancey_book, Faber11}.
We denote these degrees of freedom by $\xi$, and the frequencies and coupling constants are written as $\omega_\eta^\xi, g_\eta^\xi$ where $\eta$ identifies the degenerated components.
The effective interaction from the electron-phonon coupling can then be written as
\begin{align}
    U_{\rm ph,\eta}(\imu\nu_m) &= 
     \int \mathcal J_\eta(\omega) \frac{\omega}{(\imu\nu_m)^2 -\omega^2}
    \, \diff \omega ,
\end{align}
with the interaction density of states $\mathcal J_\eta(\omega) = \sum_\xi (g_{\eta}^{\xi})^2 \delta(\omega - \omega_\eta^\xi)$.
This is a familiar expression reminiscent of the standard Eliashberg theory \cite{Parks_book}.

\begin{figure}[t]
    \includegraphics[width = 85mm]{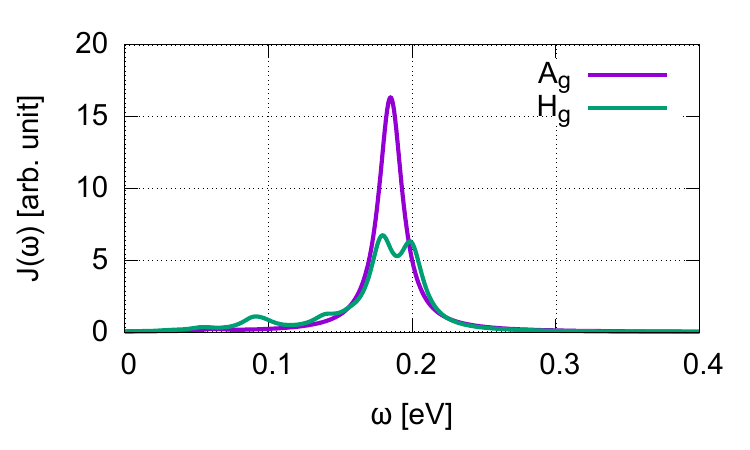}
    \caption{
        Interaction density of states $\mathcal J_\eta (\omega)$ from the contributions of two $A_{1g}$ and eight $H_g$ molecular vibrations of C$_{60}$. For the visualization, we replace delta functions by  Lorentzians with  width $0.01$eV.
    }
    \label{fig:eff_int}
\end{figure}

In Fig.~\ref{fig:eff_int}, we show the interaction density of states by using the data from the first principles calculations provided in Ref.~\cite{Faber11}.
We take the parameters for the $GW$ approximation and plot the values per  component for the one-dimensional ($A_g$) and five-degenerate representation ($H_g$).
From the figure, we see that the effective interactions for $A_{g}$ and $H_{g}$ exhibit similar peak positions and spectral height, justifying the assumption of $\eta$-independent $\omega_\eta$ and $\lambda_\eta$ used in the main text.
A similar result can also be found for the fulleride crystal \cite{Nomura15-2}.

\section{Phonon Green functions}
We consider the Green functions of $\phi=a+a^\dg=\phi^\dg$ and $p=(a-a^\dg)/\imu=p^\dg$ where the indices $(i,\eta)$ are suppressed in the Green functions for simplicity.
We define the Green functions as
\begin{align}
\begin{pmatrix}
G_{\phi\phi}(\tau) & G_{\phi p}(\tau) \\
G_{p \phi}(\tau) & G_{pp}(\tau)
\end{pmatrix}
&= - \Bigg \la \mathcal T
\begin{pmatrix}
\phi_{i\eta}(\tau) \\ 
p_{i\eta}(\tau)
\end{pmatrix}
\begin{pmatrix}
\phi_{i\eta} &  p_{i\eta}
\end{pmatrix}
\Bigg \ra
\label{eq:def_phon_A}
\end{align}
and
\begin{align}
\begin{pmatrix}
\mathscr G_{11}(\tau) & \mathscr G_{12}(\tau) \\
\mathscr G_{21}(\tau) & \mathscr G_{22}(\tau)
\end{pmatrix}
&= - \Bigg \la \mathcal T
\begin{pmatrix}
a_{i\eta}(\tau) \\ 
a_{i\eta}^\dg (\tau)
\end{pmatrix}
\begin{pmatrix}
a_{i\eta}^\dg &  a_{i\eta}
\end{pmatrix}
\Bigg \ra ,
\label{eq:def_phon_B}
\end{align}
where $A(\tau) = \epn^{\tau \mathscr H} A \epn^{-\tau \mathscr H}$ is the Heisenberg picture with imaginary time $\tau$.
We have the following  relations in the Fourier domain:
\begin{align}
    z G_{\phi\phi} &= \imu \omega_\eta G_{p\phi} = -\imu\omega_\eta G_{\phi p},
    \\
    z^2 G_{\phi\phi} &= 2\omega_\eta + \omega_\eta^2 G_{pp},
    \label{eq:phi_p_relation}
\\
\mathscr G_{11} &= \frac 1 4 (G_{\phi\phi} + 2\imu G_{p\phi} + G_{pp}),
    \\
    \mathscr G_{22} &= \frac 1 4 (G_{\phi\phi} - 2\imu G_{p\phi} + G_{pp}),
    \\
    \mathscr G_{12} &= \mathscr G_{21} = \frac  1 4 (G_{\phi\phi} - G_{pp}),
    \label{eq:offd_g}
\end{align}
where $z$ is a complex frequency and we take $z=\imu \nu_m$ for the Matsubara representation.
These relations can be used for a general Hamiltonian with the form $\mathscr H = \sum_{i\eta} \omega_\eta a_{i\eta}^\dg a_{i\eta} + \mathscr H_1 [\phi]$.
Combining Eqs.~\eqref{eq:phi_p_relation} and \eqref{eq:offd_g}, we obtain
\begin{align}
    G_{\phi\phi}^{-1} &= \frac{z^2 - \omega_\eta^2}{2\omega_\eta}
    + 2\omega_\eta G_{\phi\phi}^{-1} \mathscr G_{21},
\end{align}
where the first term on the right-hand side is the inverse of the non-interacting Green function.
Namely, the off-diagonal Green function $\mathscr G_{21} \sim \la a^\dg a^\dg\ra$, which is the difference between $G_{\phi\phi}$ and $G_{pp}$ according to Eq.~\eqref{eq:offd_g}, is proportional to the self-energy.

\begin{widetext}
\section{Self-energies in the normal state}

\subsection{Derivation of Eq.~\eqref{eq:analytic_chi}}

The calculation can be performed in a manner similar to Refs.~\cite{AGD_book, Miki21}.
The analytic continuation gives
the correlation function in the form
\begin{align}
    \chi(\imu \nu_m) 
 &= \frac{2 }{\pi} \int \diff \ep \tanh \frac{\ep}{2T} \imag G^R(\ep)
 \big[G^A(\ep-\imu\nu_m) + G^R(\ep+\imu\nu_m)\big],
\end{align}
where $\nu_m >0$ and
the superscripts $R$ and $A$ represent the retarded and advanced Green functions, respectively.
One thus finds
\begin{align}
  {\rm Re\,}  \chi^R(\omega) 
 &= \frac{2 }{\pi} \int \diff \ep \tanh \frac{\ep}{2T} \imag G^R(\ep)
 \big[{\rm Re\,} G^R(\ep-\omega) + {\rm Re\,}G^R(\ep+\omega)\big],
\\
  {\rm Im\,}  \chi^R(\omega) 
 &= - \frac{2 }{\pi} \int \diff \ep \big( \tanh \frac{\ep}{2T} - \tanh \frac{\ep-\omega}{2T} \big)
 \imag G^R(\ep)
 {\rm Im\,} G^R(\ep-\omega). 
\end{align}
We use the non-interacting Green function for a semi-circular density of states $\rho(\ep) = \frac{2}{\pi D}\sqrt{1-(\ep/D)^2}\, \theta(D-|\ep|)$,
\begin{align}
    G_0^R(\ep) 
    &= 
    \frac{2\ep}{D^2}\Big[ 1  - \sqrt{1- (D/\ep)^2} \, \theta(|\ep|-D) \Big]
     -\imu \pi \rho(\ep),
\end{align}
and thereby obtain the following retarded correlation functions at low temperatures and at low frequencies:
\begin{align}
{\rm Re\,}  \chi^R (\omega) 
&= - \frac{8 }{D^2} \int \diff \ep \rho (\ep)|\ep| + O(\omega^2)
 \simeq - \frac{32 }{3\pi D},
 \\
    {\rm Im\,}  \chi^R(\omega)  &\simeq
    - 4\pi \rho(0)^2  \omega .
\end{align}
Using the relation $\rho(0) = \frac{2}{\pi D}$ and performing the analytic continuation from the real to the imaginary axis, we find Eq.~\eqref{eq:analytic_chi}.

\subsection{Derivation of Eq.~\eqref{eq:analytic_sigma}}

We begin with the spectral representation of the self-consistent equation,
\begin{align}
   \Sigma^R (\ep) &= - \frac{1}{2\pi} \int \diff \omega \int \diff \ep' 
   \frac{\rho (\ep')\imag U_{\rm eff}^R ( \omega)}{\ep-\ep' - \omega +\imu \delta}
      \Big(   \tanh \frac{\beta \ep'}{2}+  \coth \frac{\beta \omega}{2}
 \Big).
\end{align}
The real and imaginary parts are given by
\begin{align}
  {\rm Re\,} \Sigma^R (\ep) &= - \frac{1}{2\pi} \int \diff \omega \int \diff \ep' 
   \frac{\rho (\ep')\imag U_{\rm eff}^R ( \omega)}{\ep-\ep' - \omega }
      \Big(   \tanh \frac{\beta \ep'}{2}+  \coth \frac{\beta \omega}{2} \Big),
      \\
   \imag \Sigma^R (\ep) &= \frac{1}{2} \int \diff \omega
   \rho (\ep-\omega)\imag U_{\rm eff}^R ( \omega)
      \Big(   \tanh \frac{\beta (\ep-\omega)}{2} +  \coth \frac{\beta \omega}{2}
 \Big).
\end{align}
Assuming that the phonon energy scale is small compared to the electronic energies (band width), and by using the fact that $\imag U^R$ is an odd function, we obtain the imaginary part at low energies as
\begin{align}
   \imag \Sigma^R (\max\{\ep,T\}\ll\omega_\eta) &\simeq  \rho(0) \int_0^\infty \diff \omega
  \imag U_{\rm eff}^R ( \omega)
      \Big( \frac 1 2 \tanh \frac{\beta (\ep-\omega)}{2}
      - \frac 1 2 \tanh \frac{\beta (\ep+\omega)}{2}
      +  \coth \frac{\beta \omega}{2}
 \Big)
 \\
  &\simeq  - \frac{2\pi\rho(0)}{3} \sum_\eta \frac{g_\eta^2 \gm_\eta \omega_\eta }{\tilde \omega_\eta^4} 
  \int_0^\infty \diff \omega
  \ \omega
      \Big( 
      \coth \frac{\beta \omega}{2}
      - \frac 1 2 \tanh \frac{\beta (\omega-\ep)}{2}
      - \frac 1 2 \tanh \frac{\beta (\omega-\ep)}{2}
      \Big)
      \\
      &= - \frac{\pi\rho(0)}{3}(\pi^2T^2+\ep^2) \sum_\eta \frac{g_\eta^2 \gm_\eta \omega_\eta}{\tilde \omega_\eta^4} ,
\end{align}
where we have assumed that the the dominant contribution to the integral comes from $\omega \lesssim T$
at low enough temperatures due to the Fermi distribution functions.

The real part can also be explicitly evaluated by assuming that the damping of the phonons is sufficiently small:
\begin{align}
  {\rm Re\,} \Sigma^R (\ep) 
      & =
\frac{1}{3} \sum_{\eta } \frac{g_\eta^2 \omega_\eta}{\tilde \omega_\eta} \sum_{s=\pm} \Big[ s \frac{2}{\pi D} I \Big(\frac{\ep - s\tilde \omega_\eta}{D} \Big)
      +  {\rm Re\,} G_0^R(\ep-s\tilde \omega_\eta)\Big],
      \\
      I(y) &= {\rm P}\int_{-1}^1 \diff x \frac{{\rm sgn\, }x\sqrt{1-x^2}}{y-x} 
    \simeq 2\ln \Big( \frac{\epn}{2}|y|\Big)
    \ \ ({\rm for} \ y\to 0),
\end{align}
where we have taken the low-temperature limit.
In a moderately small energy range, i.e., $\ep, \tilde \omega_\eta \ll D$, we can use the asymptotic form and obtain
\begin{align}
{\rm Re\,} \Sigma^R (\ep) 
      &=\frac{2\rho(0)}{3} \sum_{\eta } 
      \frac{g_\eta^2 \omega_\eta}{\tilde \omega_\eta}
      \bigg( \ln\left|\frac{\ep - \tilde \omega_\eta}{\ep + \tilde \omega_\eta}\right|
      + \frac{\pi \ep}{D}  \bigg).
\end{align}
Expanding the expression by $\ep$ and noting that the electronic band width is much larger than the phonon energy scales, we find the first term of Eq.~\eqref{eq:analytic_sigma}.

\end{widetext}

\end{document}